# Counting of Shortest Paths in a Cubic Grid

**Mousumi Dutt[1], Arindam Biswas[2] and Benedek Nagy[3,4]**

[1] Department of Computer Science and Engineering, St. Thomas' College of Engineering and Technology, 4, Diamond Harbour Road, Kidderpore, Kolkata-700023, West Bengal, India, mousumi.dutt@stcet.ac.in

[2] Department of Information Technology, Indian Institute of Engineering Science and Technology, Shibpur, Howrah 711103, West Bengal, India, abiswas@it.iiests.ac.in

[3] Department of Mathematics, Eastern Mediterranean University, W. Shakespeare Street, 99628 Famagusta, North Cyprus, via Mersin-10, Turkey, benedek.nagy@emu.edu.tr

[4] Institute of Mathematics and Informatics, Eszterházy Károly Catholic University, 3300 Eger, Leányka utca 4, Hungary

*Abstract: The enumeration of shortest paths in cubic grid is presented herein, which could have importance in image processing and also in the network sciences. The cubic grid considers three neighborhoods — namely,* 6-, 18- *and* 26-*neighborhood related to face connectivity, edge connectivity and vertex connectivity, respectively. The formulation for distance metrics is given.* $L_1$, $D_{18}$, *and* $L_\infty$ *are the three metrics for* 6-*neighborhood,* 18-*neighborhood and* 26-*neighborhood. The task is to count the number of minimal paths, based on given neighborhood relations, from any given point to any other, in the three-dimensional cubic grid. Based on the coordinate triplets describing the grid, the formulations for the three neighborhoods are presented in this work. The problem both of theoretical importance and has several practical aspects.*

*Keywords: cubic grid; shortest paths; combinatorics; path counting; digital distances*

## 1 Introduction

Shortest path (SP) problems have ample applications in digital geometry, which works on discrete spaces, that is, with points with integer coordinates. Based on the application, the shortest path problem can be formulated. Shortest path problems in various grids are defined based on digital distances. In the square grid, there are two classical neighbor relations defined [35] — cityblock and chessboard. The former contains horizontal and vertical movements; in chessboard motion the diagonal movements are also allowed. Consequently, two kinds of distances are defined in this grid, which are well explained in [23, 34]. In the square grid, every coordinate of a point is independent. In $n$ dimensional space, there are $n$ independent coordinates to address its elements, i.e., usually either the vertices or the hypercubes of the grid. Working in the $n$-dimensional space, the neighborhood structure of the vertices is isomorphic to the neighborhood structure of the $n$-dimensional hypercubes. The scientific field 'Geometry of Numbers' is about these grids [2, 14, 15, 18, 31]. The terms 'tiling', 'array' and 'lattice' are used approximately in the same meaning as we use 'grid' here. Counting paths as an image analysis tool has already been coined in [35], and solved with the cityblock and chessboard paths/distances in [4].

Considering non-traditional but still regular tilings, the triangular grid and the hexagonal grid have the graph-theoretic dual relation. In digital geometry, they have also been analyzed from various points of view. A connection among them and the cubic grid is established [16, 28, 31] and therefore, symmetric coordinate systems with three coordinates work nicely on these grids. The relation between square grid and hexagonal grid is explained in [38]. The three types of neighbor relation on the triangular tiling are already used in [6]. The three coordinates in this grid depend on each other [28] [30]. The digital distance of any two points, based on a fixed neighborhood criterion, is the length of a minimal-length path between the two points, where in every step along the path one moves to a neighbor point [28] [30].

The general Euclidean Shortest Path (ESP) problem is NP-hard [1] between two points amid polyhedral obstacles in the 3D space, moreover there could be exponentially many minimal path classes in single-source multiple-destination problems. A polynomial time algorithm for Euclidean Shortest Path computations, for cases where all the obstacles are convex and their number is small, is stated in [36]. The Euclidean shortest paths within a given cube-curve with arbitrary accuracy are given in [19]. ESP between two points is stated in [20-22] for 2D and 3D using rubberband algorithms. An algorithm to compute an L1-minimal path from any point to any other that lies on or above a given polyhedral terrain is presented in [24].

The discrete version of the problem is somewhat different. In graphs, a dynamic programming approach, namely the Dijkstra algorithm gives an efficient way of computing a shortest path. Digital grids can be seen as infinite graphs, where the neighbor points (pixels, voxels, etc. depending on the dimension of the used space) are connected by edges. Here, we count the number of shortest (also called minimal) paths (NSP), since there usually exist more than one shortest path depending on the conditions and on the used type of paths as a shortest path is generally not unique (similarly as in graphs). On the square grid, for any two points, a recursive formulation for counting the shortest paths between them, in cityblock, in chessboard and in octagonal approaches, is presented in [4]. It is to be noted here, that the general formulation for chessboard shortest paths, between two points was given by a recursive method based on a generating function. Herein, we also give an alternative, non-recursive formulation, based on enumerative combinatorics in Sec. 3. In [3], NSP between any pair of points, in a digital image, with respect to a particular neighbor criterion is presented, where the images are considered as matrices and thus matrix operations are used in the computation. Shortest isothetic path (cityblock) is determined between two points inside a digital object for a given grid size, in [8] [9]. Since a shortest isothetic path is usually not uniquely determined, finding the number of them is important [7]. Here, in this paper, we will discuss the path counting problem between two points whose coordinate triplets are given in cubic grid for 6- 18-, and 26-neighborhoods, i.e., $L_1$, $D_{18}$, and $L_1$ metrics respectively. The path counting problems in 3D digital geometry for the three neighborhoods are presented in [11] [12] in a different way. The formulation of path counting problem in 26-neighborhood in [11] is based on the generating function stated in [3] [4], whereas in this paper we propose it in a comprehensive and straight forward way. In [12], the formulation for path counting problem in 18-neighborhood is divided into three cases whereas the first two cases are based on the generating function stated in [3, 4] and the third one is based on induction on the length of minimal paths. The formulation for 18-neighborhood proposed here is much simpler and more definite devoid of any generating function. The computation of number of paths is more directly proposed here compared to the formulae in [11] [12]. All these formulae are proved here using combinatorial techniques.

The number of minimal paths (NSP) is related to various descriptive measures of graphs and networks including graph indices. In networking various packages may be sent in different but same length paths and in this way, NSP could be used to measure the width of the network between the given nodes. Thus, our study has not only theoretical interest, but also practical ones due to applicability both in imaging and in networking.

The paper is written in the following structure. The preliminaries are discussed in Section 2. The formulation of NSP in cubic grid for 6-*neighborhoods*, 18-*neighborhoods*, and 26-*neighborhoods* are given in Sections 3, 4, and 5 respectively. Section 6 presents concluding remarks.

# 2 Preliminaries

Based on [17], the cubic grid on $P^3$ will be denoted by $Z^3$, and defined as $Z^3 = \{(c_1, c_2, c_3) \mid c_1, c_2, c_3 \in Z\}$. Let $G$ be any set of points in $P^3$. The *Voronoi neighborhood* of $g \in G$ is defined as $N_G(g) = \{v \in P^3 \mid \forall h \in G, \|v - g\| \leq \|v - h\|\}$.

The Voronoi neighborhood of $(c_1, c_2, c_3)$ in $Z^3$ is a unit cube centered in $(c_1, c_2, c_3)$; in this way, the space is tessellated by unit cubes. When perceived as a set of points in $P^3$, $Z^3$ is referred to as a *cubic grid*. The Voronoi neighborhoods in a grid in $P^3$ are referred to, as *voxels*. Figure 1 represents the directions of the three axes in the cubic grid and the origin is also shown.

There are three widely used neighborhoods in $Z^3$. Those are 26-, 18- and 6-*neighborhood* called face-edge-vertex neighbors, face-edge neighbors and face neighbors, respectively. Let $r = (x_r, y_r, z_r) \in Z^3$ and $s_i = (x_{s_i}, y_{s_i}, z_{s_i}) \in Z^3$ be all the points fulfilling the condition max $\{|x_r - x_{s_i}|, |y_r - y_{s_i}|, |z_r - z_{s_i}|\} \leq 1$:

$$N_6(r) = \{s_i : |x_r - x_{s_i}| + |y_r - y_{s_i}| + |z_r - z_{s_i}| \leq 1\}$$

$$N_{18}(r) = \{s_i : |x_r - x_{s_i}| + |y_r - y_{s_i}| + |z_r - z_{s_i}| \leq 2\}$$

$$N_{26}(r) = \{s_i : |x_r - x_{s_i}| + |y_r - y_{s_i}| + |z_r - z_{s_i}| \leq 3\}$$

These are shown in Fig. 2. The neighbor voxels $N_6(r)$, $N_{18}(r)$, and $N_{26}(r)$ are shown in red (orange, in the figure on the right), magenta, and yellow colors. The voxels which are in 6-neighborhood of $r$, are face connected with $r$. The edge connected and vertex connected voxels are in $N_{18}(r)$, and $N_{26}(r)$ of $r$ respectively.

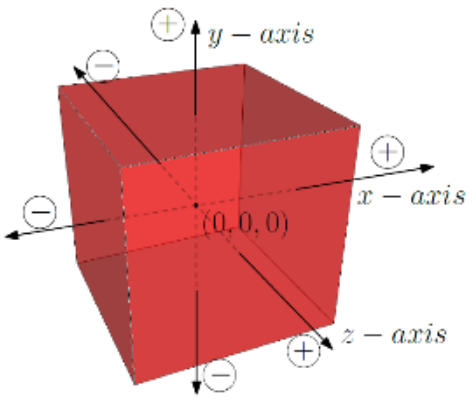

Figure 1

The origin and the directions of the three axes

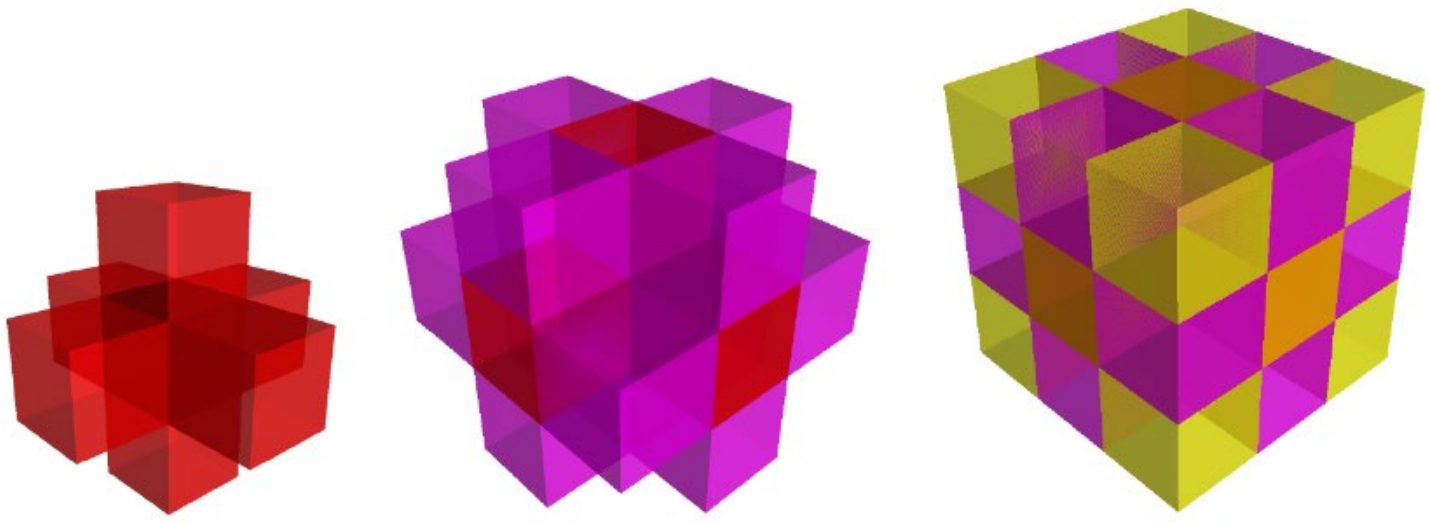

6-*neighborhood* 18-*neighborhood* 26-*neighborhood*

Figure 2

The three neighborhoods, the central cube with its 6-, 18- and 26-neighbors

Let us consider two points $q$ and $p$ in cubic grid. The problem is to find the NSP between $p$ and $q$ with a given neighbor relation. To formulate the problem, the points have to be translated such that either $p$ or $q$ be in origin (0,0,0). Let the coordinates of the points be $p = (x_p, y_p, z_p)$ and $q = (x_q, y_q, z_q)$. The coordinates of the points after translation will be $p = (x_p - x_q, y_p - y_q, z_p - z_q)$ and $q = (0,0,0)$.

We may also recall the general definition of $L_m$ distances in 3D between two points, which is given below.

$$L_m(p,q) = \left(\left|x_p - x_q\right|^m + \left|y_p - y_q\right|^m + \left|z_p - z_q\right|^m\right)^{\frac{1}{m}} \quad (1)$$

They are usually defined under the condition $m \geq 1$. The digital distances are discussed in [3, 4, 29]. We recall that the length of a minimal path from $p = (x_p, y_p, z_p)$ to $q(0,0,0)$ in cubic grid in 6-neighborhood, 18-neighborhood, and 26-neighborhood are denoted by metrics — $L_1$, $D_{18}$, and $L_\infty$ respectively, since, as it is well-known, the 6- and 26-neighborhood based distances coincide with the $L_1$ distance, and to the $L_\infty$ distance which is obtained in the limit $m \to \infty$.

$$L_1(p, q) = D_6(p, q) = (|x_p| + |y_p| + |z_p||) \quad (2)$$

$$D_{18}(p,q) = \max\left\{\max\{|x_p|, |y_p|, |z_p|\}, \left\lceil\frac{|x_p| + |y_p| + |z_p|}{2}\right\rceil\right\} \quad (3)$$

$$L_\infty(p, q) = D_{26}(p, q) = \max\{|x_p|, |y_p|, |z_p||\} \quad (4)$$

We also note that both the $L_2$ and $D_{18}$ distances are between the above mentioned “extremal” cases, i.e., both $L_1(p, q) \geq L_2(p, q) \geq L_\infty(p, q)$ and $D_6(p, q) \geq D_{18}(p, q) \geq D_{26}(p, q)$ are satisfied for any pairs of points $q$, $p \in Z^3$. Furthermore, no digital distance is known that produces $L_2$ for any pairs of points, thus to approximate the Euclidean distance by digital distances is still a hot topic both in 2D and 3D [5, 13, 25-27, 32, 33, 37].

# 3 Number of Minimal Paths in 6-Neighborhood

**Theorem 1.** The number of minimal paths from $q = (0,0,0)$ to any point $p = (i,j,k)$ in 6-neighborhood is

$$f_{6N}(i,j,k) = \frac{(|i|+|j|+|k|)!}{|i|!|j|!|k|!} \quad (5)$$

**Proof.** Without loss of generality, one may assume that the coordinates of the point $p$ are nonnegative, that is $i, j, k \geq 0$. In 3D, two points, $p'(i_1, j_1, k_1)$ and $q'(i_2, j_2, k_2)$, are in 6-neighborhood if they share a face, i.e., when the following condition holds:

$$|i_1 - i_2| + |j_1 - j_2| + |k_1 - k_2| = 1$$

Thus, since the coordinates are integers, in 6-neighborhood, in each step of a shortest path only one coordinate changes by ±1 and the other two coordinates do not change. The other coordinates of the points coincide respectively. So, the length of a shortest path between $p(i, j, k)$ and $q(0; 0; 0)$ in 6-neighborhood is $i + j + k$ (Eqn. 2), the sum of the movements along three axes. Out of total $i + j + k$ steps $i$, $j$, and $k$ steps are taken along the $x$-, $y$-, and $z$-axes respectively. Their order is arbitrary, thus the total number of arrangements for a path length of $|i| + |j| + |k|$ is given by $\frac{(|i|+|j|+|k|)!}{|i|!|j|!|k|!}$, which is the total NSP in 6-neighborhood. ■

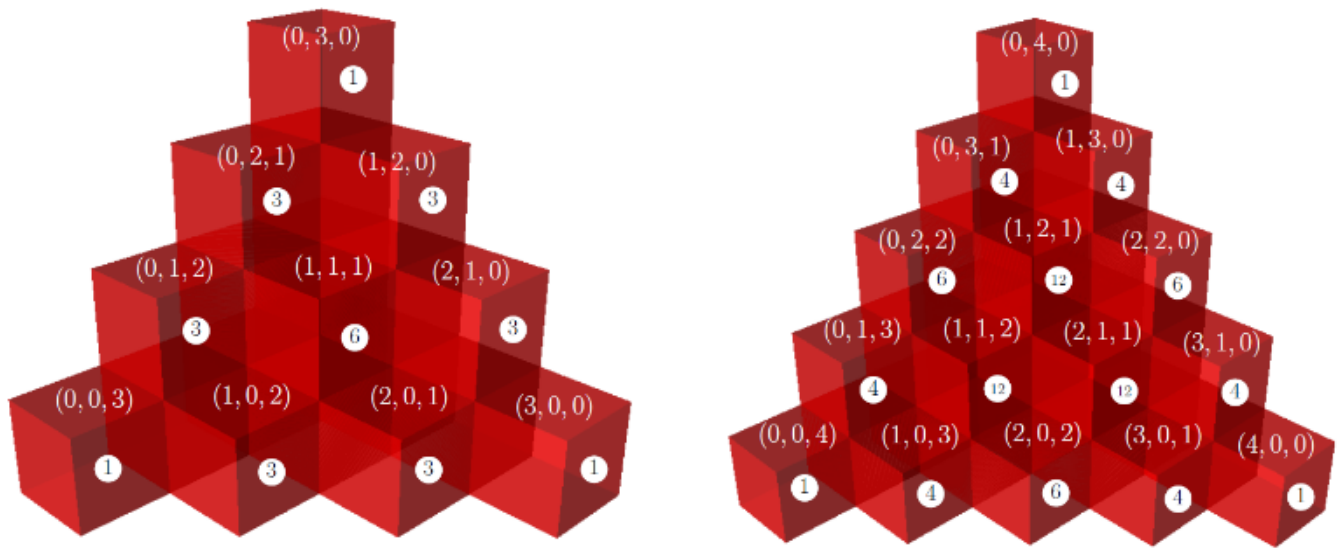

*Length=3* *Length=4*

Figure 3

In Fig. 3 the NSP of length three and four are shown for some of the points along with the coordinate triplets. Actually, these numbers are the trinomial coefficients: $\frac{(|i|+|j|+|k|)!}{|i|!|j|!|k|!}$ which could play a role in expansions like $(x + y + z)^n$, see, e.g., [4].

It is to be noted here that NSP in 6-neighborhood in cubic grid is similar for some of the coordinates where $x = 0$ or $y = 0$ or $z = 0$, with the values in 4-neighborhood in 2D, i.e., the cityblock (or $L_1$) distance in 2D, coinciding with the binomial coefficients. More formally, if the points $p(i_1, j_1, k_1)$ and $q(i_2, j_2, k_2)$ share a coordinate (e.g., $i_1 = i_2$), then their distance and NSP (in 6-neighborhood) between them can be computed in the same way as between the points analogous points in 2D with 4-neighborhood neglecting the common coordinate, i.e., between $p_{yz}(j_1, k_1)$ and $q_{yz}(j_2, k_2)$.

# 4 Number of Minimal Paths in 18-Neighborhood

Without loss of generality, we assume that $i, j, k \geq 0$. The length of a shortest path in 18-neighborhood is either maximum of $i, j, k$ or $\left\lceil \frac{i+j+k}{2} \right\rceil$ (Eqn. 3). The NSP is discussed in the following two theorems according to two cases. In our theorems $q$ will be the origin, and we are counting the paths to the point $p = (i, j, k)$, i.e., its $x$ coordinate is $i$, its $y$ coordinate is $j$ and $z$ coordinate is $k$, the coordinate axes and their directions as shown in Fig. 1.

In the next theorem, without loss of generality, we assume that $i \geq j$ and $i \geq k$, i.e., the first coordinate value of $p$ is (one of) the largest. Further, we use the variables $a$ and $b$ to denote the number of steps in specific directions made in a shortest path: whenever 2 of the coordinates are changed in a step, which is legal in this case, opposite to the previously studied $D_6$ case, there could be steps where both coordinates are increasing in a step, but also some steps where only the first coordinate is increasing and one of the other is decreasing. The variables $a$ and $b$ denote the possible numbers of such steps when the first coordinate is increasing, but either the third (variable $a$) or the second coordinate (variable $b$, resp.) is decreasing.

**Theorem 2.** Let $q = (0, 0, 0)$ and $p = (i, j, k)$ be two points such that $D_{18} = \max\{i, j, k\} = i$. Then, by using 18-neighborhood, from $q$ to $p$, the number of minimal paths is

$$f_{18N}(i,j,k) = \sum_{a=0,b=0}^{2(a+b)\le i-j-k} \frac{i!}{a!b!(k+a)!(j+b)!(i-j-k-2(a+b))!}. \quad (6)$$

where $a$ and $b$ are the number of steps in some shortest paths in right-away (positive $x$ and negative $z$) and right-bottom (positive $x$ and negative $y$) directions (based on the directions of the axes shown in Fig. 1), respectively.

**Proof.** By the symmetry of the grid, one may assume that $D_{18} = \max\{i, j, k\} = i$, i.e., there are $i$-steps from $q$ to $p$. $D_{18} = i$, implies that $i \ge j$ and $i \ge k$, moreover $i \ge j + k$ by Eqn. 3 (the cases when $D_{18} = j$ or $D_{18} = k$ are similar). In 18-neighborhood, a path can proceed through either a face-shared neighbor (change in only one coordinate value) or an edge-shared neighbor (change in any two coordinate values). In Eqn. 6, $a$ and $b$ refer to the numbers of right-away and right-bottom movements w.r.t. the positive $x$-axis (Fig. 1), respectively. Let $c = k + a$ and $d = j + b$ be the respective numbers of right and right-top movements. In a right-away movement, the path moves to the edge-shared neighbor where the $x$-coordinate increases by 1 and the $z$-coordinate decreases by 1 and in a right movement, both $x$- and $z$-coordinates increase by 1. Similarly, in a right-bottom movement $x$-coordinate increases and $y$-coordinates decreases while for a right-top movement, both the $x$ and $y$ coordinates increase. The sum of movements cannot be more than $i$: $a + b + c + d \le i$, i.e., $2(a + b) \le i - j - k$. The right-away and the right movements as well as the right-top and the right-bottom movements have some limits. When a number of right-away movements are there, the right movements will be $c = k + a$ such that the decrease of $z$-coordinate in a right-away moves is compensated by the increase of the $z$-coordinate in $k + a$ right moves in order that the destination point has $z$-coordinate as $k$. Note that in each move the $x$-coordinate always increases by 1. Similarly, $b$ number of right-bottom movements implies $d = j + b$ number of right-top movements. Otherwise, it will not be possible to reach the destination in $i$ steps. Apart from right-away, right, right-top, and right-bottom moves, there can be movements in $x$-direction only. For a given $a$, $b$, $c$, and $d$ (i.e., right-away, right-bottom, right, and right-top respectively) steps, there are $i - (a + b + c + d) = i - j - k - 2(a + b)$ number of steps in the positive $x$-direction to face neighbor. Thus, for a given $a$, $b$, $c$, and $d$ combination, the total number of arrangements for a shortest path of length $i$ is given by $\frac{i!}{a!b!(k+a)!(j+b)!(i-j-k-2(a+b))!}$. For different values of $a$ and $b$, values of $c$ and $d$ are computed satisfying the condition that $a + b + c + d \le i$. Thus, total NSP is the summation over the different possible combinations of $a$, $b$, $c$, and $d$ values, and is given by Eqn. 6. ■

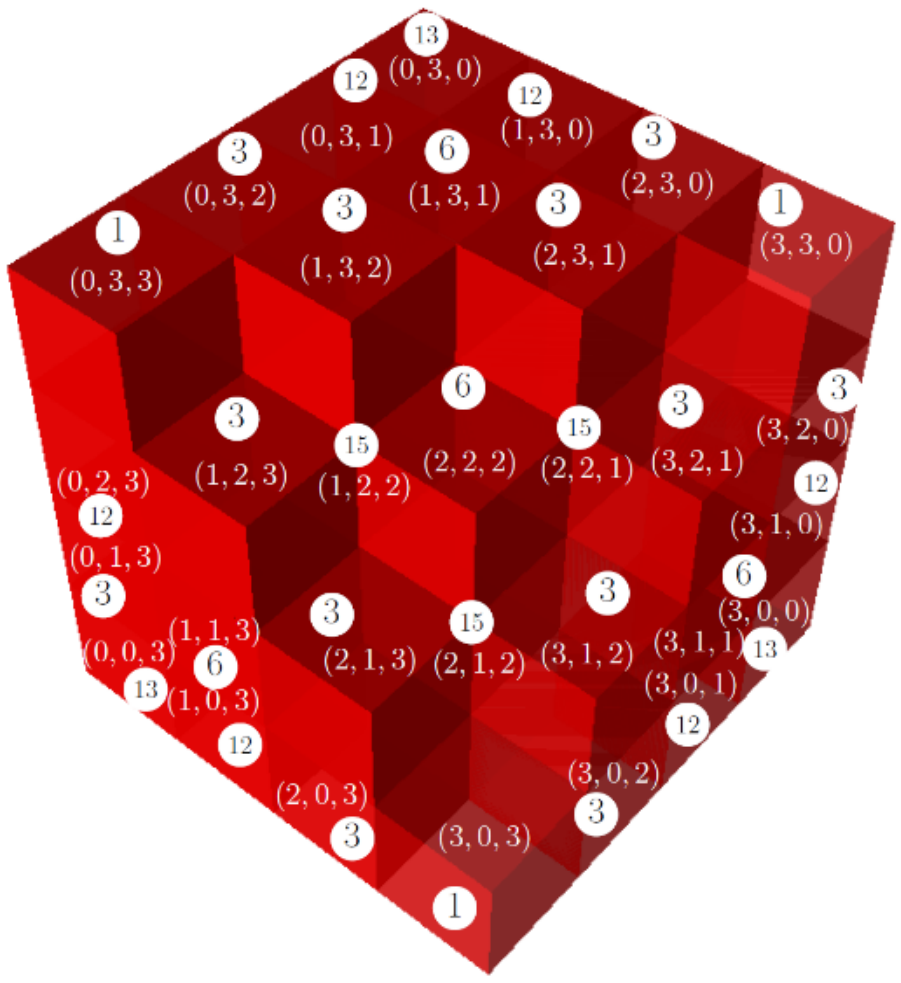

*Length=3*

Figure 4
The NSP (given inside the white circle) from origin to other points (coordinates are shown in parentheses) in 18-neighborhoods for path length three

The NSP from $q(0, 0, 0)$ to $p$ where the length of path is three, are given in Fig. 4. It is to be noted that when $D_{18}$ = max $\{i, j, k\} = j$ or $k$, the above formula (Eqn. 6) will change accordingly. The number of paths from (0, 0, 0) to (0, 3, 0) is 13 and that to (0, 3, 1) is 12.

**Theorem 3.** The number of minimal paths from $q$ = (0, 0, 0) to any point $p = (i, j, k)$ in 18-neighborhood when $D_{18} = \left\lceil \frac{i+j+k}{2} \right\rceil = \tau$ is

$$f_{18N}(i,j,k) = \begin{cases} \dfrac{\tau!}{(\tau-i)!(\tau-j)!(\tau-k)!}, & \text{when } (i+j+k) \bmod 2 = 0 \\ \dfrac{\tau!((\tau-i)(\tau-j)+(\tau-j)(\tau-k)+(\tau-k)(\tau-i))}{(\tau-i)!(\tau-j)!(\tau-k)!}, & \text{when } (i+j+k) \bmod 2 = 1 \end{cases}. \tag{7}$$

**Proof.** At each step in 18-neighborhood, at most two coordinates can increase by one and the number of steps is τ. When $i + j + k$ is even, as τ divides $i + j + k$ with the quotient 2 which implies that at each step always two (distinct) coordinates will increase. Therefore, in a shortest path from $q(0, 0, 0)$ to $p(i, j, k)$ of length τ, the number of steps in both $y$- and $z$-directions is $\tau - i$ , in both $x$- and $z$-directions is $\tau - j$, and in both $x$- and $y$-directions is $\tau - k$. Thus, the number of possible arrangements, i.e., the NSP is $\frac{\tau!}{(\tau-i)!(\tau-j)!(\tau-k)!}$

When $i + j + k$ is odd, τ divides $i + j + k + 1$ and the quotient is 2 as $D_{18} = \left\lceil\frac{i+j+k}{2}\right\rceil = \tau$, it implies that for $\tau - 1$ steps two coordinates will increase and in the rest one step only one of the three coordinates will increase (let it be called a singular step) giving rise to the following cases:

***Singular step in x-direction***: A shortest path has one singular step in $x$-direction. Thus, there are rest $\tau - 1$ steps, where at each step there are movements in two directions. The number of steps when there are movements in $x$- and $y$-directions in each step is $(\tau - 1) - k$, in $x$- and $z$-direction is $(\tau - 1) - j$, and in $y$- and $z$-direction is $(\tau - 1) - (i - 1) = \mathrm{L} - i$. So, the number of possible shortest paths with singular $x$-direction is $\frac{\tau!}{(\tau-i)!((\tau-1)-j)!((\tau-1)-k)!} = \frac{\tau!(\tau-j)(\tau-k)}{(\tau-i)!(\tau-j)!(\tau-k)!}$. When $j > i + k$ and $\tau = j$ or $k > i + j$ and $\tau = k$ the singular step in $x$-direction will never occur.

***Singular step in y-direction***: There will be one singular step in $y$-direction. Here, the number of steps in $x$- and $y$-direction is $(\tau - 1) - k$, in $x$- and $z$-direction is $(\tau - 1) - (j - 1) = \tau - j$, and in $y$- and $z$-direction is $(\tau - 1) - \mathrm{i}$, giving the number of possible shortest path with singular $y$-direction as $\frac{\tau!(\tau-i)(\tau-k)}{(\tau-i)!(\tau-j)!(\tau-k)!}$. When $i > j + k$ and $\tau = i$ or $k > i + j$ and $\tau = k$ the singular step in $y$-direction will never occur.

***Singular step in z-direction***: One of the steps will be in the $z$-direction. The number of steps in $x$- and $y$-direction is $(\tau - 1) - (k - 1) = \tau - k$, in $x$- and $z$-direction is $(\tau - 1) - j$, and in $y$- and $z$-direction is $(\tau - 1) - i$. Thus, the number of possible shortest paths with singular $z$-direction is given by $\frac{\tau!(\tau-i)(\tau-j)}{(\tau-i)!(\tau-j)!(\tau-k)!}$. When $j > i + k$ and $\tau = j$ or $i > j + k$ and $\tau = i$ the singular step in $z$-direction will never occur.

Hence, the total NSP when $i + j + k$ is odd is given by $\frac{\tau!(\tau-j)(\tau-k)}{(\tau-i)!(\tau-j)!(\tau-k)!} + \frac{\tau!(\tau-i)(\tau-k)}{(\tau-i)!(\tau-j)!(\tau-k)!} + \frac{\tau!(\tau-i)(\tau-j)}{(\tau-i)!(\tau-j)!(\tau-k)!} = \frac{\tau!\left((\tau-i)(\tau-j)+(\tau-j)(\tau-k)+(\tau-k)(\tau-i)\right)}{(\tau-i)!(\tau-j)!(\tau-k)!}$. ■

The NSP for path length three is shown in Fig. 4. The NSP from (0, 0, 0) to (2, 2, 2) is 6 where $i + j + k$ is even and that to (1, 2, 2) is 15 where $i + j + k$ is odd. The number of paths from (0, 0, 0) to (0, 3, 2) satisfy both the equations stated in Theorem 2 and 3 and that from (0, 0, 0) to (2, 3, 1) also satisfy both the equations (see Fig. 4). To compute NSP between (0, 0, 0) and (9, 5, 4), the formula stated in Theorem 2 and 3 (here, $i + j + k$ is even) both are applicable and produce same result, 630. Similarly, to find NSP between (0, 0, 0) and (9, 4, 4), the formula stated in Theorem 2 and 3 (here, $i + j + k$ is odd) both yield same result 630. Remember that the distance $D_{18}$ is computed as the maximum of a set. In some cases, it may happen that there are more maximal elements of this set, and thus, both Theorem 2 and 3 can be applied to count NSP. In these cases, they must give the same value, as we state formally in the following.

**Corollary 1.** The number of minimal paths from $q = (0, 0, 0)$ to any point $p = (i, j, k)$ in 18-neighborhood when $D_{18} = \left\lceil \frac{i+j+k}{2} \right\rceil = \max\{i, j, k\} = \tau = i$, $f_{18N}(i, j, k)$ is as follows.

$$f_{18N}(i,j,k) = \sum_{a=0,b=0}^{2(a+b)\le i-j-k} \frac{i!}{a!b!(k+a)!(j+b)!(i-j-k-2(a+b))!} = \begin{cases} \frac{\tau!}{(\tau-i)!(\tau-j)!(\tau-k)!}, & \text{when } (i+j+k) \bmod 2 = 0 \\ \frac{\tau!((\tau-i)(\tau-j)+(\tau-j)(\tau-k)+(\tau-k)(\tau-i))}{(\tau-i)!(\tau-j)!(\tau-k)!}, & \text{when } (i+j+k) \bmod 2 = 1 \end{cases} \quad (8)$$

**Proof.** The proof can be done mathematically in two parts when $i + j + k$ is even and when it is odd. Let $\tau = \frac{i+j+k}{2} = i$, i.e., $i + j + k$ is even. Thus, $i - j - k = 0$. Putting $i - j - k = 0$, in Eqn. 6 (see Theorem 2) we get, $f_{18N}(i,j,k) = \sum_{a=0,b=0}^{2(a+b)\le 0} \frac{i!}{a!b!(k+a)!(j+b)!(-2(a+b))!}$, as there is only one possibility for the values of $a$ and $b$, i.e., $a = b = 0$ since $2(a+b) \le i - j - k = 0$. By putting these values, we get $\frac{i!}{j!k!}$. Since, $\tau = i$, $i - j = k$ and $i - k = j$. Putting these values in the first expression of Eqn. 7 (see Theorem 3) when $i + j + k$ is even, we get $\frac{i!}{j!k!}$. Hence proved.

For the second part, when $i + j + k$ is odd, $\tau = \frac{i+j+k+1}{2} = i$. Thus, $i - j - k = 1$. Putting $i - j - k = 1$ in Eqn. 6 (see Theorem 2) we get, $f_{18N}(i,j,k) = \sum_{a=0,b=0}^{2(a+b)\le 1} \frac{i!}{a!b!(k+a)!(j+b)!(1-2(a+b))!}$, as there is one possibility for the values of $a$ and $b$, i.e., $a = b = 0$ since $2(a+b) \le i - j - k = 1$. By putting these values, we get $\frac{i!}{j!k!}$. Now, $\tau - i = 0$, $\tau - j = k + 1$, $\tau - k = j + 1$. Thus, $\frac{\tau!((\tau-i)(\tau-j)+(\tau-j)(\tau-k)+(\tau-k)(\tau-i))}{(\tau-i)!(\tau-j)!(\tau-k)!} = \frac{i!(0+(j+1)(k+1)+0)}{0!(j+1)!(k+1)!} = \frac{i!}{j!k!}$. Hence proved. ■

Similarly, the above-mentioned equation (Eqn. 8) can be proved when $\tau = j$ or $k$.

# 5 Number of Minimal Paths in 26-Neighborhood

The formulation for NSP in 26-neighborhood in cubic grid depends on NSP in 8-neighborhood in 2D. This is exactly the chessboard distance in 2D [35]. The NSP in 8-neighborhood in 2D had been proposed by Das [3, 4] with recurrence relations, in this paper we show a shorter direct proof with combinatorial tools (Eqn. 9). We count NSP from the origin to a point $p(i, j)$. Without loss of generality, we may assume that the absolute value of the first coordinate of $p$ is not less than the absolute value of its second coordinate, i.e., $|i| \ge |j|$. Similarly, as in Theorem 2, we use a variable, here $b$, to denote the possible number of steps that in which both coordinates are changed, the first is changed in the direction of $p$ from $q$, while the second one is changed in opposite way. Based on that we formulate the result in the next theorem, and one may see the formal details in the proof.

**Theorem 4.** Let $q = (0, 0)$ be the origin, and let point $p = (i, j)$ be such that their distance is $i$. Then, in 2D with 8-neighborhood, the number of minimal paths from $q$ to the point $p$ is given by

$$f_8(i,j) = \sum_{b=0}^{2b\leq|i|-|j|} \frac{|i|!}{b!\,(|j|+b)!\,(|i|-|j|-2b)!} \quad \text{where } |i| \geq |j| \tag{9}$$

**Proof.** The length of a shortest path between $p(i, j)$ and $q(0, 0)$ in 8-neighborhood is max $\{|i|, |j|\}$. By the symmetry of the grid, we show the proof for the case $0 \leq j \leq i$, in this case the distance is $i$. With respect to the positive $x$-direction, let $b$ be the number of moves along right-bottom diagonal in the shortest path where $x$-coordinate increases by 1 and $y$-coordinate decreases by 1, and $d = |j| + b$ be the number of moves along right-top diagonal in the shortest path where the $x$- and $y$-coordinates increases by 1. A shortest path involves $i$ steps, out of which if there are $b$ right-bottom moves then $d = j + b$ moves only in right-top direction, hence, the number of paths is given by $\frac{|i|!}{b!(j+b)!(i-j-2b)!}$. It may be noted here that $b + d \leq i$, i.e., $i - j - 2b$ as the total number of moves cannot be more than $i$. By summing over the different possible combinations of $b$ and $d$, the total number of shortest paths is given by $f_8(i,j) = \sum_{b=0}^{2b\leq i-j} \frac{|i|!}{b!(j+b)!(i-j-2b)!}$. ■

The number of paths from $q(0, 0)$ to other points in 8-neighborhood in 2D are shown in Fig. 5. Figure 6 shows two examples of all possible paths from a source to destination. For a particular path (shown in red) among all possible shortest paths, the $l$ and $r$ values are given for ease of understanding. The formulation for $|j| > |i|$, is just reverse (exchanging $i$ with $j$) of the above equation (Eq. 9). The values appearing in 8-neighborhood in 2D are also present in 26-neighborhood of cubic grid counting the number of shortest paths from the origin to $p(i, j, k)$ if $i = j \geq k$ or $j = k \geq i$ or $i = k \geq j$ (see Fig. 5 and Fig. 7).

Now we are ready to state our last result. Without loss of generality, we count the NSP from the origin to a point $p = (i, j, k)$ such that their distance is $i$, i.e., $|i| \geq |j|$ and $|i| \geq |k|$. The variables $a$ and $b$ are used to count those steps where the $y$ and the $z$ coordinates are changing not to the direction of $j$ and $k$, respectively, i.e., they are decreasing if the appropriate coordinate of $p$ is nonnegative.

**Theorem 5.** The number of minimal paths from $q = (0, 0, 0)$ to any point $p = (i, j, k)$ in 26-neighborhood is

$$f_{26N}(i,j,k) = \sum_{b=0}^{2b\le|i|-|j|} \frac{|i|!}{b!\,(|j|+b)!\,(|i|-|j|-2b)!} \times \sum_{a=0}^{2a\le|i|-|k|} \frac{|i|!}{a!\,(|k|+a)!\,(|i|-|k|-2a)!} \tag{10}$$

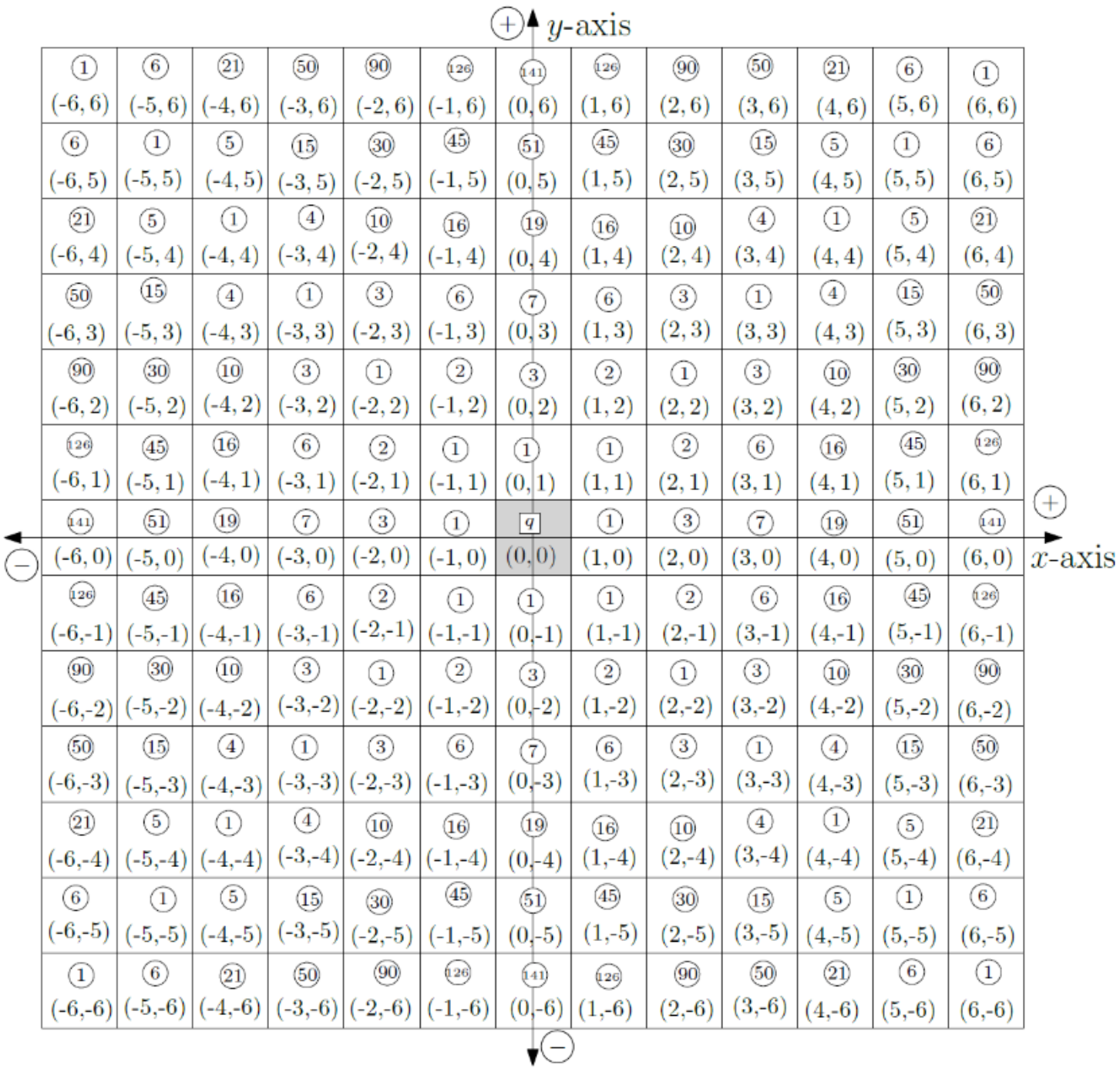

Figure 5
The NSP from origin to other points in 8-neighborhood in 2D. (The coordinate pairs are written in parentheses and the corresponding NSP values are also mentioned.)

**Proof.** The length of a minimal path between $p(i, j, k)$ and $q(0, 0, 0)$ in 26-neighborhood is max $\{|i|, |j|, |k|\} = i$ (given in Eqn. 4) (say). In each step of a shortest path in 26-neighborhood, at most three coordinates can change. A shortest path in 26-neighborhood is a combination of a shortest path in $xy$-plane, from $q(0, 0, 0)$ to $p_{xy}(i, j, 0)$ and a shortest path in $xz$-plane, from $q(0, 0, 0)$ to $p_{xz}(i, 0, k)$. With each shortest path in $xy$-plane, each shortest path in $xz$-plane is combined to get the total NSP in 3D. Thus, the NSP in 26N is given by $f(i, j) \times f(i, k)$ where $f(i, j)$ and $f(i, k)$ are NSPs in $xy$- and $xz$-planes respectively (Theorem 4). Thus,

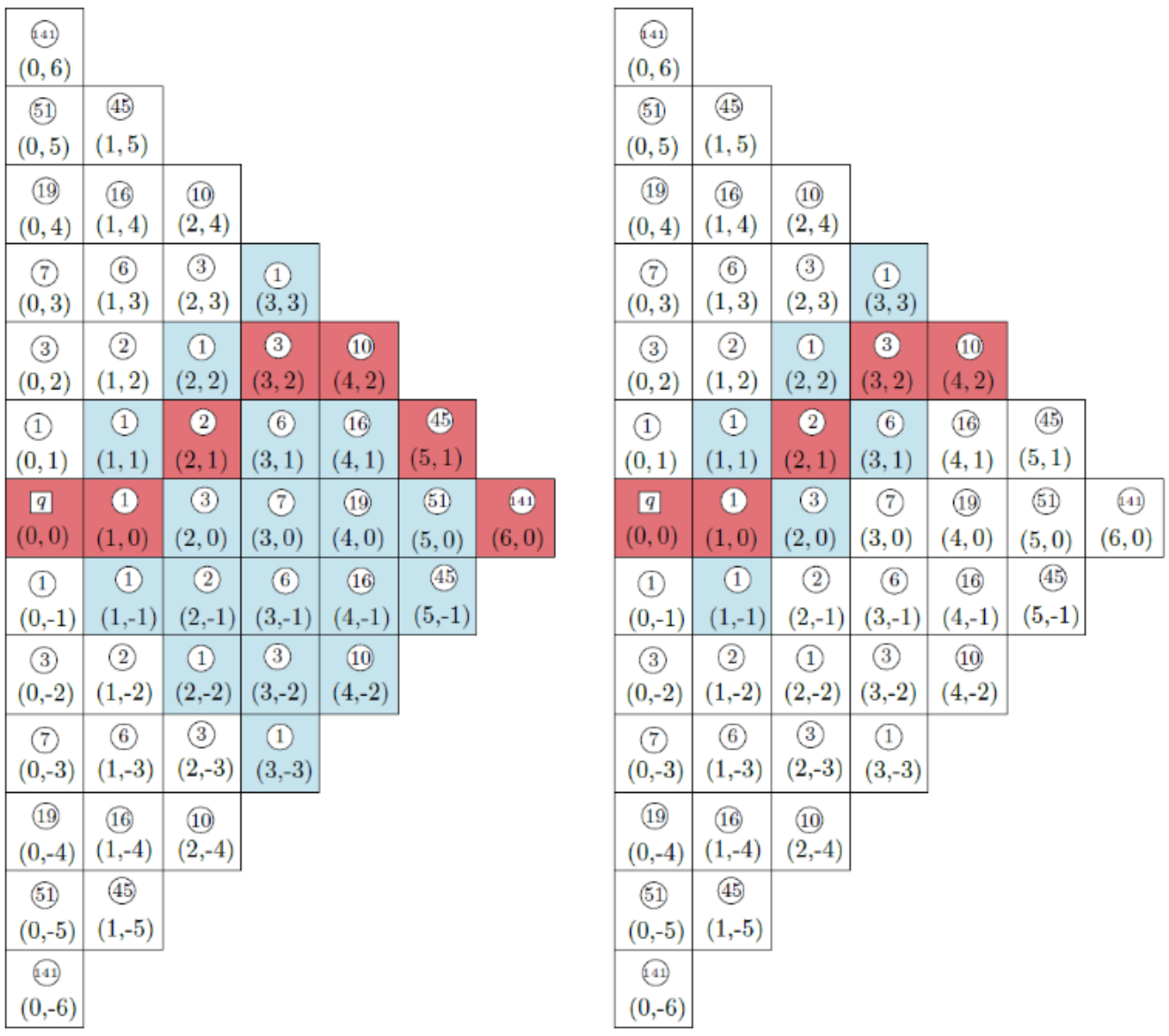

Figure 6
The NSP between the origin and other points in 8-neighborhood in 2D. The shaded portion shows the cells covered by all possible paths between two points out of which one path is shown by red color where $|j| \geq |i|$. The path in left figure has $b = 3$ and $d = |j| + b = 3$ and that of right figure is $b = 0$ and $d = |j| + b = 2$.

$$f_{26N}(i,j,k) = \sum_{b=0}^{2b\leq|i|-|j|} \frac{|i|!}{b!(|j|+b)!(|i|-|j|-2b)!} \times \sum_{a=0}^{2a\leq|i|-|k|} \frac{|i|!}{a!(|k|+a)!(|i|-|k|-2a)!}$$

Where, $a$ and $b$ indicate the numbers of steps in right-bottom projected directions (moves that simultaneously increasing the first and decreasing the second coordinates, i.e., in positive $x$- and in negative $y$-directions, the $z$-directions might be arbitrary, i.e., ±1 or +0 for these moves) and right-away projected directions (moves in positive $x$- and negative $z$-directions, by increasing the first and decreasing the third coordinate, while the second coordinate might change by ±1 or not in these moves), respectively, if $j$ and $k$ are nonnegative. ■

From the Equation 10, the formulation for NSP between two points for $|j| \geq |k|$, $|i|$ and $|k| \geq |i|$, $|j|$ can be derived similarly. The NSP of length three between $q(0, 0, 0)$ and some other points are shown in Fig. 7. Figure 8 shows an example path from $q(0, 0, 0)$ to $p(7, 4, 2)$ which has 2 right-away movements with corresponding 4 right movements and 1 right-bottom with corresponding 5 right-top movements.

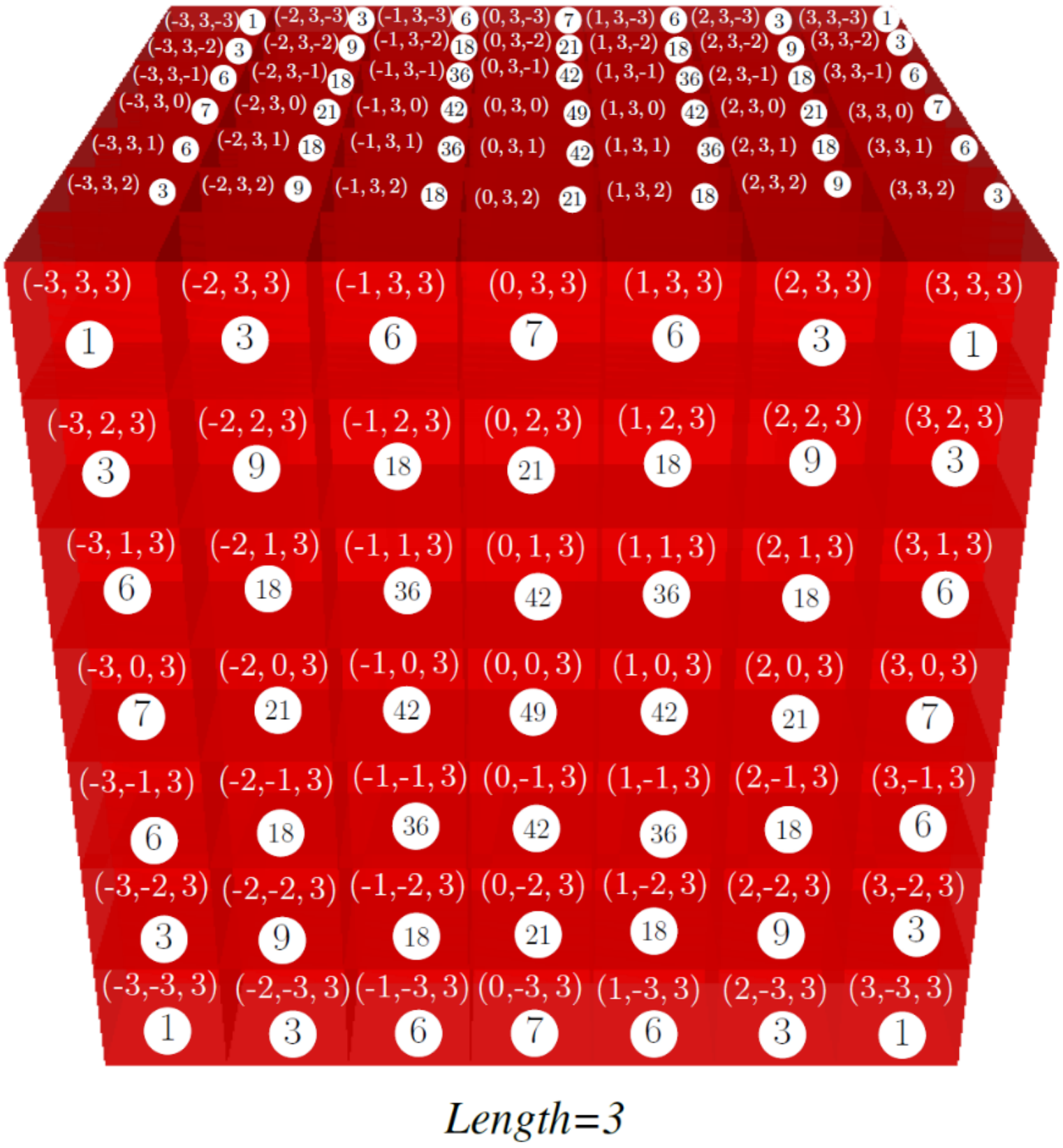

Figure 7

The NSP from origin to other points in 26-neighborhoods for *Length* = 3. (The coordinate triplets are written in parentheses and the corresponding NSP are also mentioned.) Observe that the results are not only symmetric, but they are according to a multiplication table, by Eqn. 10, where the elements of the border rows and columns are specified by the formula for the 2D $L_\infty$ distance, i.e., the chessboard distance (Eqn. 9). Obviously, the diagonals contain the squares of the numbers shown at the borders.

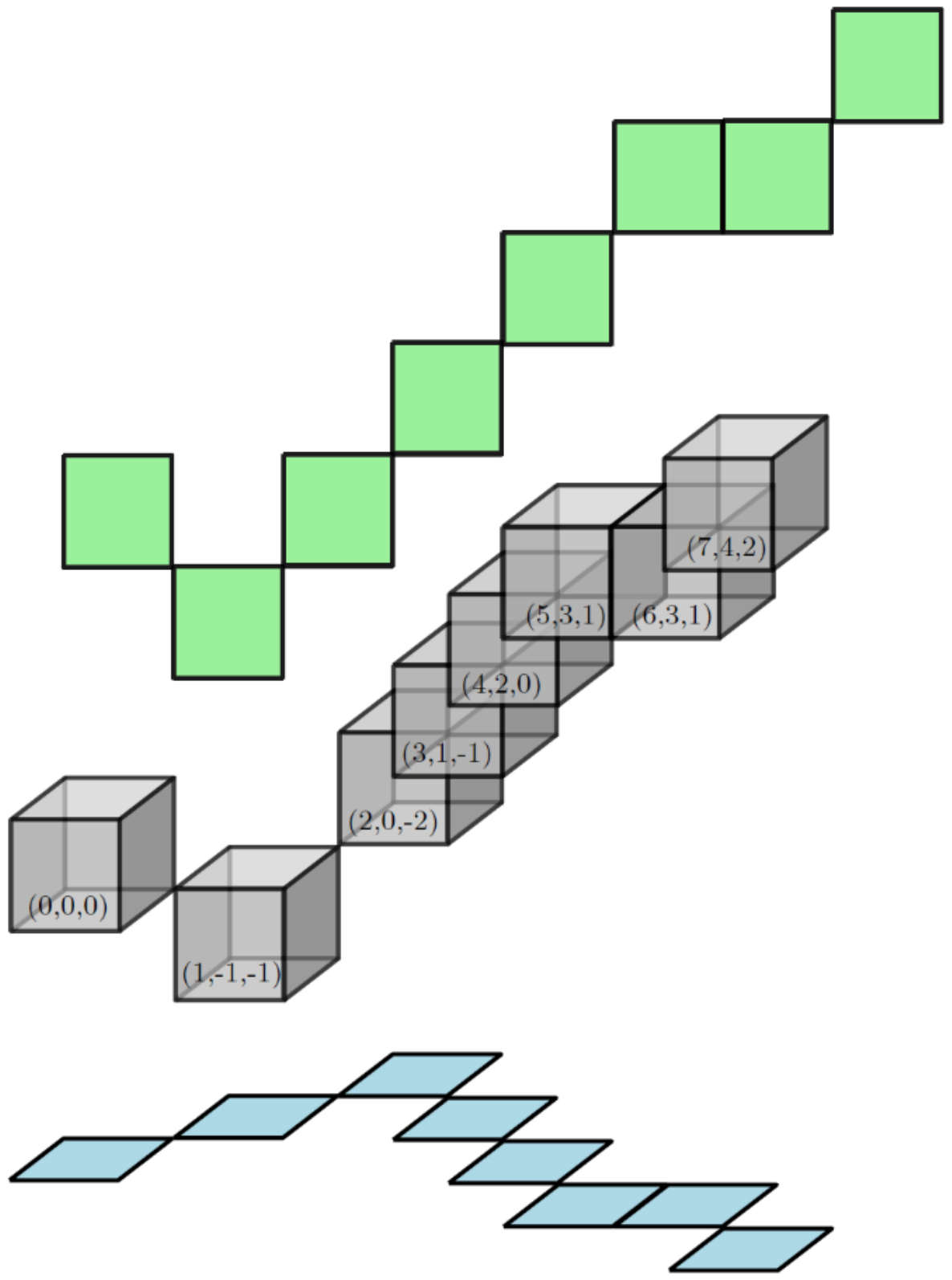

Figure 8

One of the shortest paths of length 7 from (0, 0, 0) to (7, 4, 2) is shown and correspondingly the values of $l$, $r$, $b$, and $t$ are 2, 4, 1, and 5 respectively. The projection of the paths in $xy$-plane is shown at back in green color and that in $xz$-plane in blue color at the bottom.

## Conclusions

The shortest path problem has various applications in several fields, especially in image processing and image analysis. Digital distances are some of the important features in this regard and many studies have already been presented. In this paper, extending the results of Das [4] from 2D to 3D, using $L_1$, $D_{18}$ and $L_\infty$ distances, the number of minimal paths (NSP) between any point pair in the cubic grid are presented for 6-, 18- and 26-neighborhood where the coordinate triplets of the two points are provided. It is also to be noted that the formulation for the NSP in 8-neighborhood in 2D is stated in this paper using combinatorial tool. In future, NSP problem in cubic grid can be extended for general orthogonal polyhedron. A 3D object can be represented by 3D orthogonal polyhedron. The critical points at different parts of 3D orthogonal polyhedrons need to be identified and the numbers of paths, between all such pairs of points, are important features for the shape analysis of 3D objects. Similar to the methods of path counting were applied in 2D images in [35].

**Acknowledgements**

The paper has an earlier version archived and can be found as [10].